\documentclass[twocolumn,prl]{revtex4}
\usepackage{natbib}
\usepackage{definitions}
\usepackage{localdefs}
\usepackage{amsmath}
\usepackage{amssymb}
\usepackage{graphicx}
\usepackage{amsbsy}
\usepackage{bm}

\begin{document}
\date{\today}

\title{Hydrodynamic crystals: collective dynamics of regular arrays of
spherical particles in a parallel-wall channel }

\author{M. Baron}
\affiliation{Department of Mechanical Engineering, Yale University,
P.O. Box 20-8286, New Haven, CT 06520}
\author{J.\ B{\l}awzdziewicz}
\affiliation{Department of Mechanical Engineering, Yale University,
P.O. Box 20-8286, New Haven, CT 06520}
\author{E.\ Wajnryb}
\altaffiliation{On leave from IPPT Warsaw, Poland.}
\affiliation{Department of Mechanical Engineering, Yale University,
P.O. Box 20-8286, New Haven, CT 06520}

\pacs{xxx}

\begin{abstract}

Simulations of over $10^3$ hydrodynamically coupled solid spheres are
performed to investigate collective motion of linear trains and
regular square arrays of particles suspended in a fluid bounded by two
parallel walls.  Our novel accelerated Stokesian-dynamics algorithm
relies on simplifications associated with the Hele--Shaw asymptotic
far-field form of the flow scattered by the particles.  The
simulations reveal propagation of particle-displacement waves,
deformation and rearrangements of a particle lattice, propagation of
dislocation defects in ordered arrays, and long-lasting coexistence of
ordered and disordered regions.

\end{abstract}

\maketitle

Long-range hydrodynamic interactions between solid particles suspended
in a fluid result in complex collective dynamic phenomena, such as
development of ordered arrays of magnetically driven rotors placed on
a liquid interface
\cite{magnetic_rotors}
and formation of time-dependent patterns in a system of particles immersed in
a vibrated fluid \cite{Voth-Bigger-Buckley-Losert-Brenner-Stone-Gollub:2002}.
Collective behavior due to the hydrodynamic coupling also occurs in biological
systems.  A striking example is spontaneous formation of vortical arrays of
self-propelled sperm cells confined to an interface
\cite{Riedel-Kruse-Howard:2005}.  Hydrodynamic coupling also plays an
essential role in the synchronization of cilia beating and development of
collective waves in cilia arrays in small swimming organisms
\cite{cilia}.

In confined multiphase systems, the collective particle behavior is strongly
influenced by  bounding walls affecting the fluid motion.  According to
recent studies
\cite{Cui-Diamant-Lin-Rice:2004,%
Bhattacharya-Blawzdziewicz-Wajnryb,%
Beatus-Tlusty-Bar_Ziv:2006},  
hydrodynamic confinement effects are especially significant in
parallel-wall channels of width comparable to the particle size.
Lateral motion of a particle in such a channel produces fluid backflow
that is involved in numerous dynamical phenomena. It enhances relative
particle motion in confined quasi-2D-suspensions
\cite{%
Cui-Diamant-Lin-Rice:2004,%
Bhattacharya-Blawzdziewicz-Wajnryb%
},
considerably increases transverse hydrodynamic resistance for elongated rigid
arrays of spheres moving parallel to the channel walls
\cite{%
Bhattacharya-Blawzdziewicz-Wajnryb%
},
and governs propagation of particle-displacement waves
\cite{Beatus-Tlusty-Bar_Ziv:2006} in linear arrays of drops in a
microfluidic channel.  We show that the fluid backflow resulting from
particle motion is also responsible for pattern formation occurring in
2D hydrodynamic crystals (i.e.\ regular particle arrays that are
hydrodynamically driven).

\begin{figure}[b]

\includegraphics*{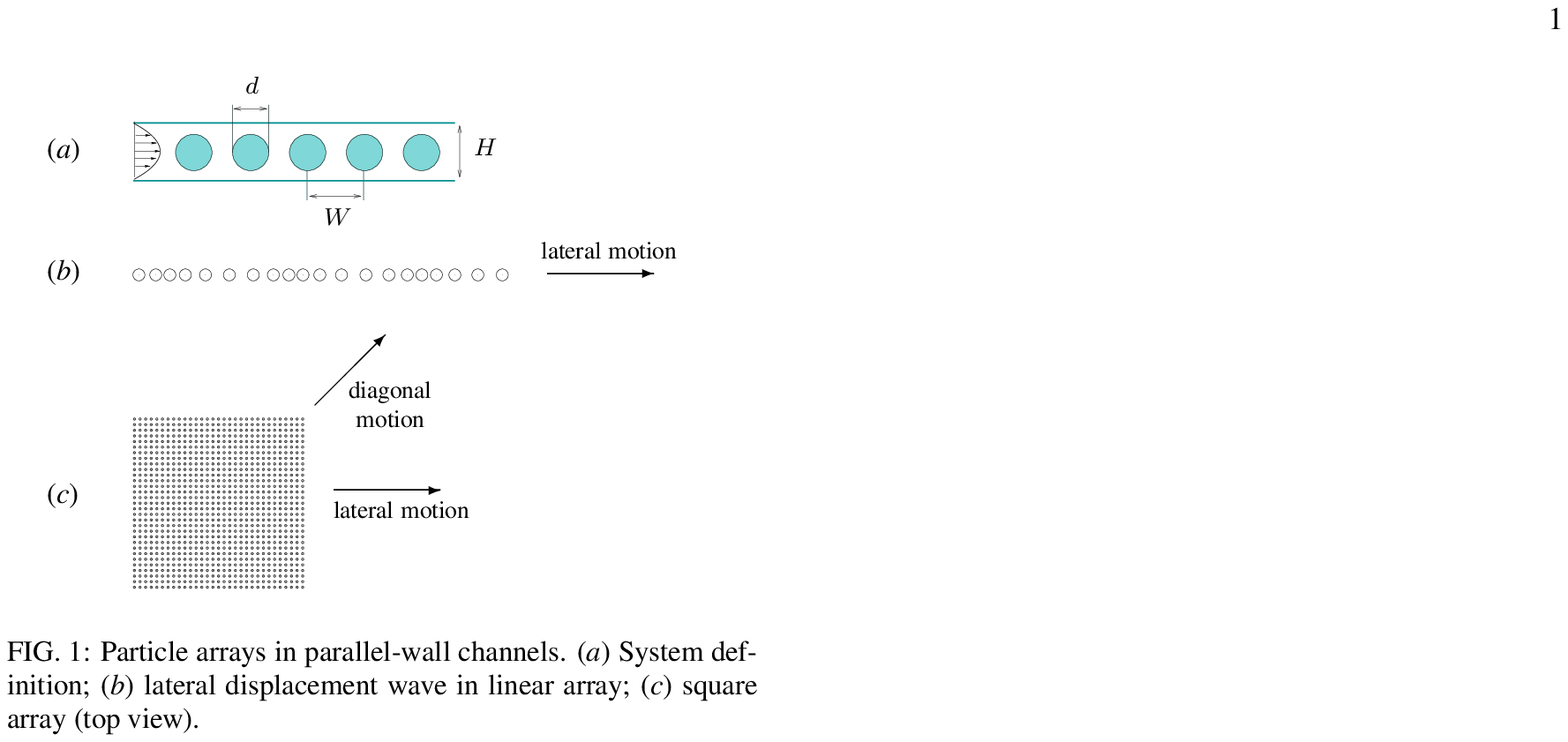}
\caption{
Particle arrays in parallel-wall channels. \subfig{a} System
definition; \subfig{b} lateral displacement wave in linear array;
\subfig{c} square array (top view).
}

\label{definition figure}
\end{figure}

In this Letter we present a numerical study of the dynamics of 1D and
2D regular arrays of hydrodynamically coupled spherical particles in
parallel-wall channels (cf.\ configurations shown in Fig.\
\ref{definition figure}).  We investigate propagation of displacement
waves in linear arrays. In large square 2D arrays we report emergence
of striking patterns, such as rearrangements of particle lattice,
dislocation defects, and coexistence of ordered and disordered
domains.  We show that these patterns occur as a result of macroscopic
deformation of a regular particle lattice, and we propose a
macroscopic theory describing shape evolution of the arrays.

\begin{figure}

\includegraphics*{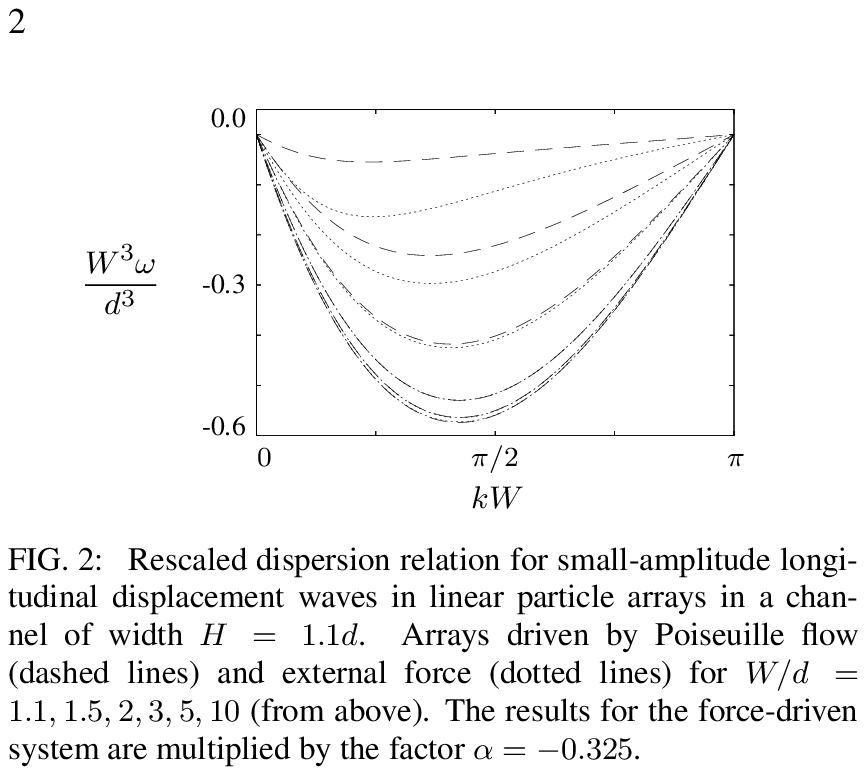}
\caption{\small{
Rescaled dispersion relation for small-amplitude longitudinal displacement
waves in linear particle arrays in a channel of width $H=1.1d$.  Arrays driven
by Poiseuille flow (dashed lines) and external force (dotted lines) for
$\particleSeparation/d=1.1, 1.5, 2, 3, 5, 10$ (from above).  The results for
the force-driven system are multiplied by the factor $\alpha=-0.325$.
}}
\label{Dispersion relation plot}
\end{figure}

\begin{figure}
\includegraphics*{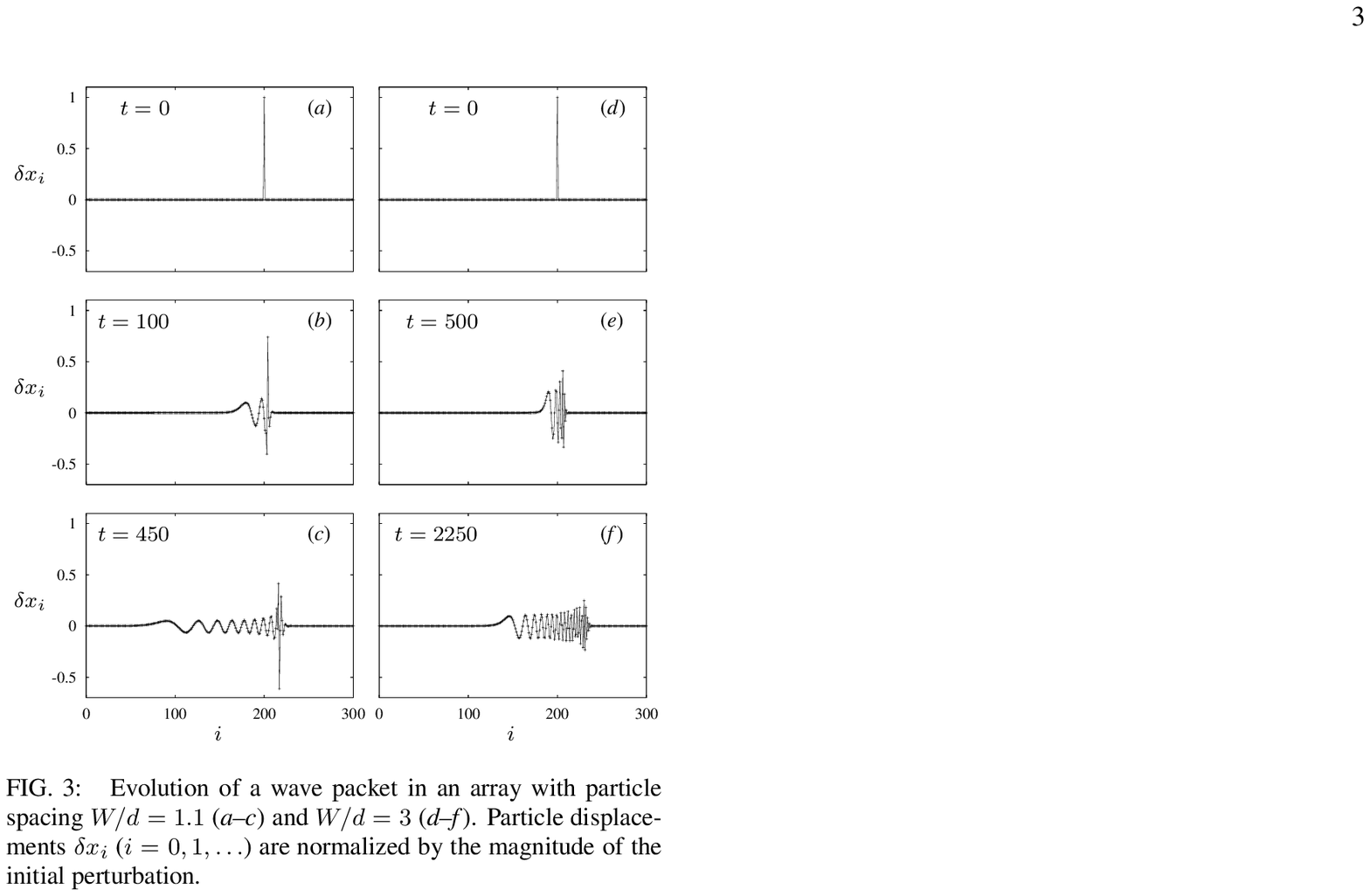}
\caption{\small{
Evolution of a wave packet in an array with particle spacing
$\particleSeparation/d=1.1$ \subfig{a--c} and $\particleSeparation/d=3$
\subfig{d--f}.  Particle displacements $\normalizedDisplacement{i}$
($i=0,1,\ldots$) are normalized by the magnitude of the initial perturbation.
}}
\label{wave packets}
\end{figure}

Our simulations are performed using a novel accelerated
Stokesian-dynamics algorithm to follow evolution of about $10^3$
particles.  Potential applications of our new algorithm include
studies of collective motion of self-propelled particles (e.g.\
bacterial colonies) in liquid films, modeling suspension flows in slit
pores, and investigations of dynamics of macromolecules (e.g.\ DNA or
polymer chains) in microfluidic channels.  Our acceleration technique
can also be used in boundary-integral algorithms for studying dynamics
of deformable particles in confined geometry.

Our numerical technique relies on simplifications associated with the
far-field asymptotics of the flow scattered from the particles.  Far
from a particle, the scattered flow in a parallel-wall channel assumes
the Hele--Shaw form, i.e.\ it tends to a 2D parabolic flow that is
driven by a harmonic pressure distribution
\cite{Bhattacharya-Blawzdziewicz-Wajnryb}.  In our new approach we
expand the flow scattered by the particles into a carefully chosen
fundamental set of Stokes flows. Close to a particle the basis flows
form a complete set of solutions of Stokes equations in 3D space. In
the far-field domain $\rho\gg H$ (where $\rho$ is the lateral distance
from the particle, and $H$ is the wall separation) these flows either
exponentially tend to zero or to Hele--Shaw flow driven by a 2D
pressure multipole.  The expansion of the flow field into the new set
of basis fields (obtained by an orthogonal transformation from the
fields used in
\cite{Bhattacharya-Blawzdziewicz-Wajnryb}) 
yields a sparse system of linear equations, which can be efficiently
solved using iterative sparse-matrix-manipulation techniques.
Moreover, since the far-field flow is uniquely determined by the
harmonic pressure distribution, well-developed acceleration techniques
for Laplace equation can be applied to further increase numerical
efficiency.  The simulations discussed below show that our
algorithm is efficient and highly accurate in both the far-field and
the near-field regimes.  The calculations also reveal surprisingly
rich collective particle dynamics.

Figures \ref{Dispersion relation plot} and \ref{wave packets} present
our results for propagation of particle displacement waves in an
infinite train of equally spaced particles positioned along a line in
the midplane of a channel slightly wider than the particle size.  The
particle array is driven either by Poiseuille flow [cf. Fig.\
\ref{definition figure}\subfig{a}] or by a constant external lateral
force.  We focus on the longitudinal waves, where the particle
displacements $\displacement{i}$ from the reference positions $x_i=i
\particleSeparation$ ($i=0,1,2\ldots$) on a regular lattice with
spacing $\particleSeparation$ occur along the array [cf.\ Fig.\
\ref{definition figure}\subfig{b}].

Figure\ \ref{Dispersion relation plot} shows the dispersion relation
$\omega=\omega(k)$ for harmonic displacement waves
$\displacement{i}=\mbox{$\epsilon\sin(kx_i-\omega t)$}$ in arrays with
different interparticle spacing.  Here $k$ is the wave number,
$\omega$ is the wave frequency measured in the reference frame moving
with the particles, and $\epsilon\ll1$ is the wave amplitude.  The
time and frequency are normalized by the time $\tauW$ in which an
isolated particle in a channel moves by the diameter $d$.  The shape
of the dispersion curves is reflected in the evolution of wave packets
depicted in Fig.\ \ref{wave packets}.  For small interparticle spacing
the maximum frequency is shifted towards smaller wave vectors, because
the lubrication forces hinder the relative particle motion.  Hence,
there is a long-wave tail in the wave packet shown in Fig.\ \ref{wave
packets}\subfig{c}.

\begin{figure}

\includegraphics*{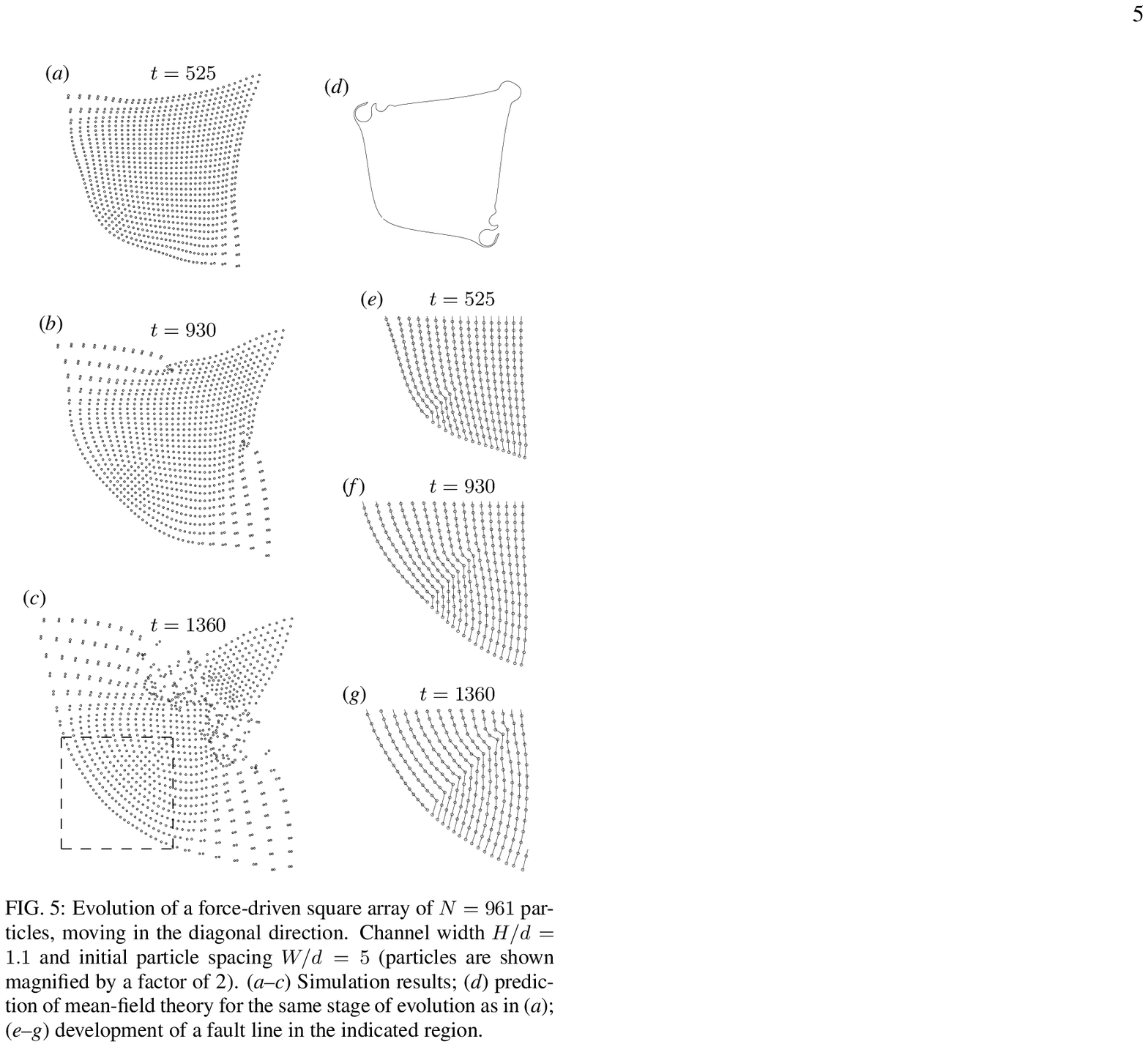}

\caption{
Evolution of a force-driven square array of $N=961$ particles, moving
in the diagonal direction.  Channel width $H/d=1.1$ and initial
particle spacing $\particleSeparation/d=5$ (particles are shown
magnified by a factor of 2).  \subfig{a--c} Simulation results;
\subfig{d} prediction of mean-field theory for the same stage of
evolution as in \subfig{a}; \subfig{e--g} development of a fault line
in the indicated region.
}
\label{airplane 31}
\end{figure}

In Fig.\ \ref{Dispersion relation plot} the frequency $\omega$ is
plotted rescaled by a factor $(\particleSeparation/d)^3$ to emphasize
the universal behavior of the system for large values of
$\particleSeparation$.  In addition, the results for the force-driven
train are multiplied by a constant negative factor $\alpha$.  We find
that for $\particleSeparation/d\gtrsim5$ all rescaled results fall
onto a single asymptotic master curve.  In the regime
$\particleSeparation/d\approx \mbox{2--3}$ the dispersion relations 
significantly deviate from the master curve, but the scaled
results for the flow- and force-driven arrays are still nearly
identical.  Since the scale factor $\alpha$ is negative, this behavior
indicates that for moderate and large interparticle distances the
relative particle motion in an array driven by an external flow is
equivalent to the relative motion in a force-driven array moving in
the {\it opposite\/} direction.

\begin{figure}

\includegraphics*{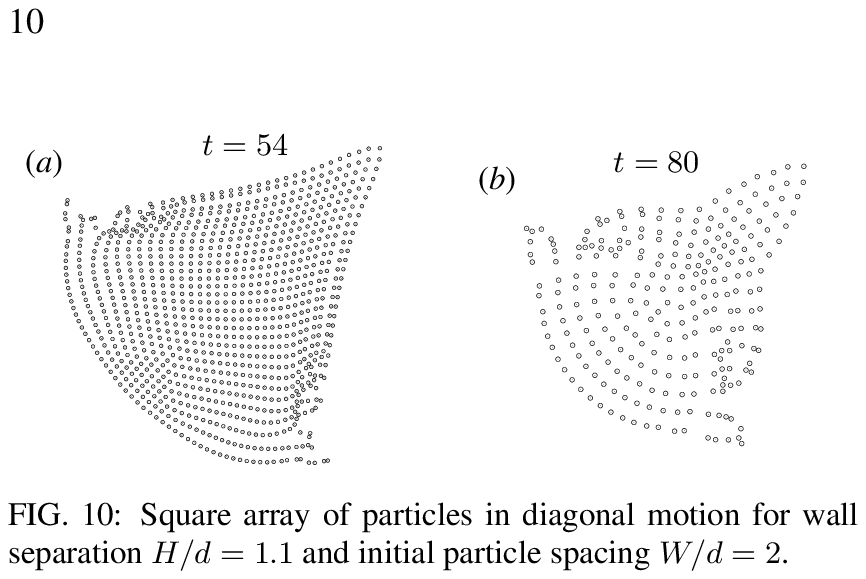}
\caption{%
Square arrays of \subfig{a} $N=961$ and \subfig{b} $N=256$ particles in
diagonal motion. Channel width $H/d=1.1$ and initial particle spacing
\hbox{$\particleSeparation/d=2$}.}
\label{Two airplanes 16}
\end{figure}

The above features of the system dynamics result from the far-field
behavior of the flow field scattered by the particles.  In the
far-field regime an isolated particle subject to Poiseuille flow or
external force produces the same Hele--Shaw flow $\HeleShawDipole$
driven by the two-dimensional dipolar pressure
$\pressureDipole\sim\cos\phi/\rho$ (where $\phi$ is the polar angle
measured from the direction of the external forcing)
\cite{%
Cui-Diamant-Lin-Rice:2004,%
Bhattacharya-Blawzdziewicz-Wajnryb%
}.
For $\particleSeparation/d\gtrsim5$ the single-scattering
approximation corresponding to the superposition of dipolar fields
$\HeleShawDipole$ adequately describes the system dynamics (and thus
the rescaled results in Fig.\ \ref{Dispersion relation plot} follow a
single master curve).  In the regime $\particleSeparation/d\approx
\mbox{2--3}$ evaluation of a single flow reflection is insufficient;
nevertheless, the results for flow- and force-driven arrays can be
rescaled onto each other.  This is because the first reflection in the
multiple-scattering sequence for the two systems is nearly identical
(apart from a rescaling factor), owing to the exponential approach of
the flow field to the asymptotic Hele--Shaw form $\HeleShawDipole$.
With matching initial reflections, the whole multiple-scattering
sequences for systems with different forcing coincide, and the
relative particle motion is thus the same.  This argument is valid not
only for linear arrays but also for other horizontal particle
arrangements (such as the 2D arrays shown in Figs.\ \ref{airplane
31}--\ref{Two airplanes 16}). Moreover, similar reasoning applies to
different kinds of forcing, including Marangoni and electrophoretic
forces used to control particle positions in microfluidic devices.

\begin{figure}[b]
\includegraphics*{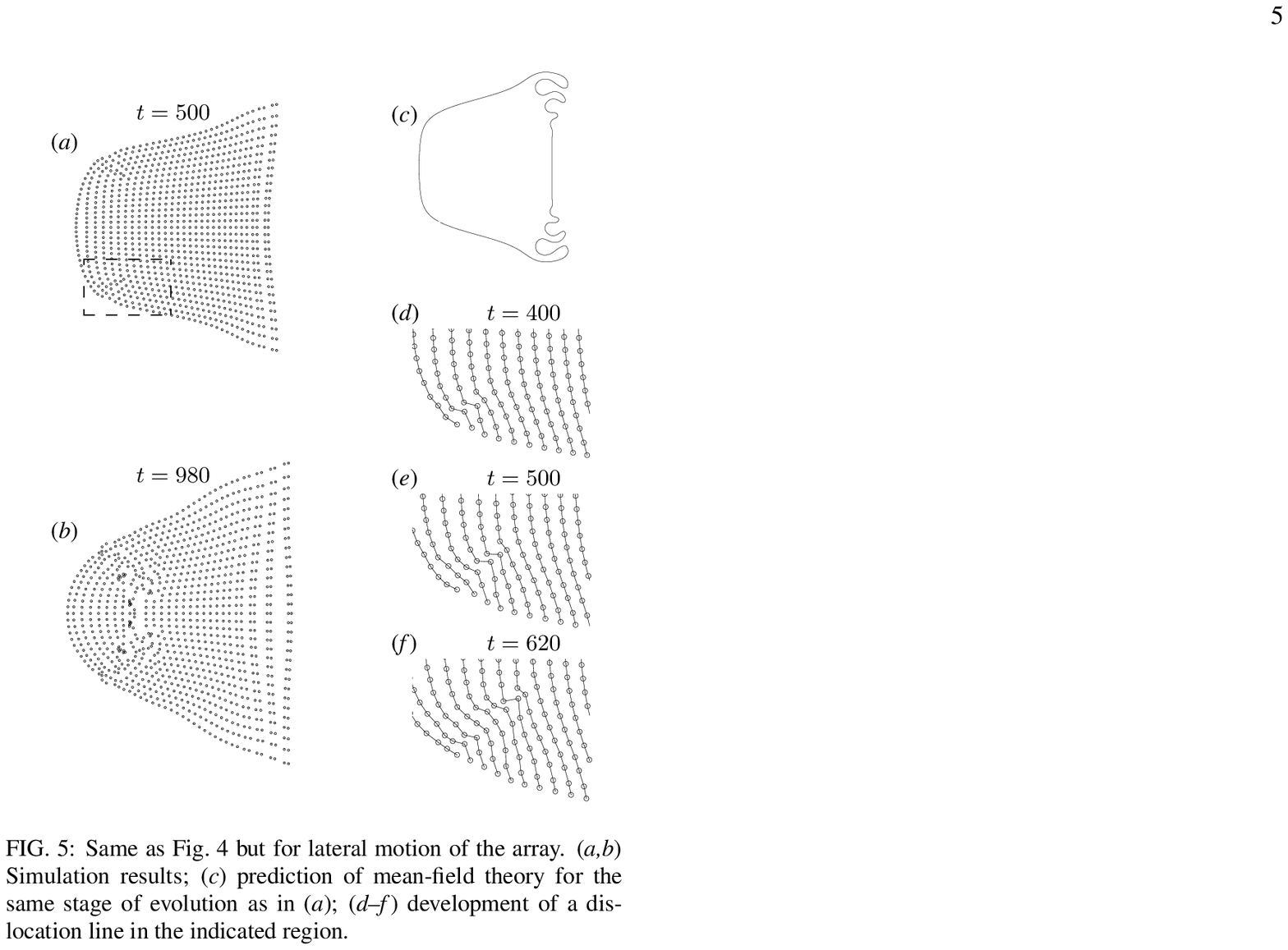}

\caption{Same as Fig.\ \protect\ref{airplane 31} but for lateral
motion of the array.  \subfig{a,b} Simulation results; \subfig{c}
prediction of mean-field theory for the same stage of evolution as in
\subfig{a}; \subfig{d--f} development of a dislocation line in the
indicated region.}
\label{carpet 31}
\end{figure}

Rich collective phenomena revealed by our simulations of 2D
hydrodynamic crystals are illustrated in Figs.\ \ref{airplane
31}--\ref{Two airplanes 16}. Figure \ref{airplane 31} presents the
evolution of a regular square array [cf. Fig.\ \ref{definition
figure}\subfig{c}] of about $10^3$ particles undergoing diagonal
motion produced by a constant force acting on all the particles. The
initial particle spacing is within the far-field asymptotic regime,
$\particleSeparation/d=5$.  Figure \ref{carpet 31} shows corresponding
results for lateral motion of the array.

Our simulations demonstrate that at short times a deforming square
array retains its initial particle ordering [cf. Fig.~\ref{airplane
31}\subfig{a}].  Subsequently, the system develops some striking
structural features.  Several rows of particle pairs separate from the
main body of the array, forming a shape similar to airplane wings.
The front part of the array has an approximately hexagonal particle
ordering, and the middle part retains the square ordering.  The rear
part [marked region in Fig.\ \ref{airplane 31}\subfig{c}] has a square
particle arrangement but with a different orientation than the
original one. The blowup in Fig.\ \ref{airplane 31}\subfig{e--g} shows
that the particle rearrangement involves discontinuous particle
displacements along a ``fault line'' at the symmetry axis of the
array.  A similar dislocation event (but without lattice
reorientation) is observed in an array in the lateral motion
(cf. Fig.\ \ref{carpet 31}).  There also occur instabilities
responsible for emergence of disordered domains of chaotic particle
motion.  In the diagonal motion [Fig. \ref{airplane
31}\subfig{b,c}] the instability starts at the junction between the
wings and the body of the array.  For the lateral motion, the
order-disorder transition occurs when the dislocation lines approach
the center of the array [Fig. \ref{carpet 31}\subfig{b}].

Figures \ref{airplane 31} and \ref{carpet 31} demonstrate that the
ordered regions can withstand large macroscopic deformations and
random perturbations originating from the disordered
domains. Moreover, the ordered crystalline domains can rearrange and
heal themselves along fault or dislocation lines.  The strong
propensity to maintain the ordered structure results from the dipolar
hydrodynamic interactions of neighboring particles.  An array also
undergoes a macroscopic deformation resulting from the combined
long-range effect of the dipolar flow fields produced by individual
particles.

For low-density arrays (i.e.\ for $W/d\gg1$) the macroscopic flow that
causes the deformation can be determined from the flow field produced
by a uniform distribution of pressure dipoles $\pressureDipole$
induced in the array. In general, the macroscopic deformation can be
described using the effective 2D transport equations for suspension
flow in a parallel-wall channel,
$\volumeFlux=-\mobilityVV \pressureDrop+\mobilityVP\forceDensity$, and
$\particleFlux=-\mobilityPV \pressureDrop+\mobilityPP\forceDensity$.
Here $\volumeFlux$ is the suspension velocity averaged across the
channel, $\particleFlux$ is the particle flux, $\forceDensity$ denotes
the density of the lateral force acting on the particles,
$\macroscopicPressure$ is the macroscopic pressure, and $\nu_\alpha$,
$\mu_\alpha$ ($\alpha=p,f$) are linear transport coefficients.  The
suspension velocity and suspension flux satisfy the continuity
equations $\bnabla\bcdot\volumeFlux=0$ and $\partial \bar\rho/\partial
t=-\bnabla\bcdot\particleFlux$, where $\bar\rho$ is the suspension
density.

The macroscopic deformation of an array, evaluated in the
uniform-dipolar-moment approximation, is shown in Figs.\ \ref{airplane
31}\subfig{d} and \ref{carpet 31}\subfig{c}.  The results indicate
that our macroscopic description reproduces the overall shape of the
arrays for moderate times (i.e.\ before the complex structural
features develop).  

We note that the macroscopic equations predict fingering instabilities
near the array corners.  In low-density arrays (cf.\ Figs.\
\ref{airplane 31} and \ref{carpet 31}) such instabilities are
suppressed due to the array ``stiffness'' associated with its ordered
structure.  However, for denser arrays (cf. Fig.\ \ref{Two airplanes
16}) the macroscopic deforming forces are sufficiently strong to
destabilize the tips of the array wings, in agreement with our
macroscopic theory.  The results in Fig.\ \ref{Two airplanes 16}
indicate that the lengthscale for the fingering instability in
particle arrays is determined by the particle lattice.  In our
effective medium theory there is no intrinsic lengthscale, so the size
of the fingers in Figs.\ \ \ref{airplane 31}\subfig{d} and \ref{carpet
31}\subfig{c} is set by the initial condition (i.e.\ a square with
rounded corners).

The patterns we observe in 2D hydrodynamic crystals have analogies in
other athermal systems.  For example, dislocations and chaotic
dynamics develop in arrays of convective cells in a fluid undergoing
Benard convection \cite{Young-Riecke:2003} and in vibrated granular
media \cite{Aranson-Tsimring:2006}.  Our system has a number of
interesting distinctive features.  First, the pattern formation occurs
in the linear Stokes-flow regime, and the nonlinearity stems entirely
from the position-dependence of the multiparticle mobility matrix.
Next, the dipolar hydrodynamic interactions that maintain particle
ordering are non-isotropic (causing, e.g., lattice reorientation).
Finally, the dipolar flow $\HeleShawDipole$ not only maintains the
ordered structure on the local level but also produces the macroscopic
deformation of the array, leading to lattice instabilities.

Regular particle arrays can be assembled using holographic optical
tweezers \cite{tweezers}, so the collective dynamic phenomena revealed
by our study should be accessible experimentally.  The effect of
hydrodynamic coupling on the motion of regular particle arrays could
also be observed in flow-driven 2D colloidal crystals.  The
equivalence of the relative particle motion in systems with different
forcing can be used to separately control the relative particle
positions and the position of the center of mass of an array.

\acknowledgments{This work was supported by NSF CAREER grant
CBET-0348175 and MNiSW grant N501 020 32/1994.}

\bibliographystyle{apsrev} 
\bibliography{/home/jerzy/BIB/jbib}

\end{document}